\documentclass[twoside,english,aps, prx, superscriptaddress, amsfonts, amssymb, amsmath, a4paper, reprint]{revtex4-1}
\usepackage{lmodern}

\usepackage[LGR,T1]{fontenc}
\usepackage[utf8]{inputenc}
\usepackage{color}
\usepackage{babel}
\usepackage{booktabs}
\usepackage{textcomp}
\usepackage{amsmath}
\usepackage{amssymb}
\usepackage{stmaryrd}
\usepackage{graphicx}
\usepackage[unicode=true,
 bookmarks=false,
 breaklinks=true,pdfborder={0 0 0},pdfborderstyle={},backref=false,colorlinks=true]
 {hyperref}
\hypersetup{pdftitle={Relaxation Dynamics of Spin Qubits in Quantum Dots},
 pdfauthor={Dan Cogan}}

\makeatletter

\ProvideTextCommand{\~}{LGR}[1]{\char126#1}

\providecommand{\tabularnewline}{\\}

\newcommand{\ket}[1]{\left|#1\right\rangle }
\newcommand{\bra}[1]{\left\langle #1\right|}
\newcommand{\im}{\operatorname{Im}}
\newcommand{\Tr}{\operatorname{Tr}}

\makeatother

\begin{document}

\title{Depolarization of Electronic Spin Qubits Confined in Semiconductor
Quantum Dots. }

\author{Dan Cogan}

\affiliation{The Physics Department and the Solid State Institute, Technion–Israel
Institute of Technology, 3200003 Haifa, Israel}

\author{Oded Kenneth}

\affiliation{The Physics Department and the Solid State Institute, Technion–Israel
Institute of Technology, 3200003 Haifa, Israel}

\author{Netanel H. Lindner}

\affiliation{The Physics Department and the Solid State Institute, Technion–Israel
Institute of Technology, 3200003 Haifa, Israel}

\author{Giora Peniakov}

\affiliation{The Physics Department and the Solid State Institute, Technion–Israel
Institute of Technology, 3200003 Haifa, Israel}

\author{Caspar Hopfmann}

\affiliation{The Physics Department and the Solid State Institute, Technion–Israel
Institute of Technology, 3200003 Haifa, Israel}

\author{Dan Dalacu}

\affiliation{National Research Council of Canada, Ottawa, Ontario, Canada K1A
0R6}

\author{Philip J. Poole }

\affiliation{National Research Council of Canada, Ottawa, Ontario, Canada K1A
0R6}

\author{Pawel Hawrylak }

\affiliation{Physics Department, University of Ottawa, ON, Canada ON Canada K1N
6N5}

\author{David Gershoni}
\email{dg@physics.technion.ac.il}

\selectlanguage{english}%

\affiliation{The Physics Department and the Solid State Institute, Technion–Israel
Institute of Technology, 3200003 Haifa, Israel}
\begin{abstract}
Quantum dots are arguably the best interface between matter spin qubits
and flying photonic qubits. Using quantum dot devices to produce joint
spin-photonic states requires the electronic spin qubits to be stored
for extended times. Therefore, the study of the coherence of spins
of various quantum dot confined charge carriers is important both
scientifically and technologically. In this study we report on spin
relaxation measurements performed on five different forms of electronic
spin qubits confined in the very same quantum dot. In particular,
we use all optical techniques to measure the spin relaxation of the
confined heavy hole and that of the dark exciton – a long lived electron-heavy
hole pair with parallel spins. Our measured results for the spin relaxation
of the electron, the heavy-hole, the dark exciton, the negative and
the positive trions, in the absence of externally applied magnetic
field, are in agreement with a central spin theory which attributes
the dephasing of the carriers' spin to their hyperfine interactions
with the nuclear spins of the atoms forming the quantum dots. We demonstrate
that the heavy hole dephases much slower than the electron. We also
show, both experimentally and theoretically, that the dark exciton
dephases slower than the heavy hole, due to the electron-hole exchange
interaction, which partially protects its spin state from dephasing. 
\end{abstract}
\maketitle
\global\long\def\ket#1{\left|#1\right\rangle }
\global\long\def\bra#1{\left\langle #1\right|}
\global\long\def\im{\operatorname{Im}}
\global\long\def\Tr{\operatorname{Tr}}

\section{Introduction}

The electronic spin in semiconductor nanostructures can often be described
as an isolated physical two level system. As such it has long been
considered an excellent qubit with great potential to be used in future
quantum information processing based technologies~\citep{Loss1998,spinsasqubits,Kimble2008}.
Moreover, semiconductor nanostructures, which confine single electrons,
are easily integrated into electronic and optical devices and circuits,
which dovetail with the contemporary semiconductor based electro-optic
technology. Therefore, many efforts have been devoted recently to
demonstrate that various forms of the electronic spin in semiconductor
nanostructures and in particular in quantum dots (QDs) can be initiated
and controlled with relatively high fidelities, using optics and electronics
means~\citep{Berezovsky_2008,Press_2008,Ramsay_2008,Michler2017}.
An important advantage of semiconductor electronic spin qubits, which
are anchored to the device, is their strong interaction with photons,
which can be used as flying qubits to communicate quantum information
to remote locations \citep{Imamog_lu_1999,Gao2012,Greve2012,Schaibley2013}.
These advantages have been recently used for instance, to demonstrate
that a QD confined electronic spin, can be used as an entangler for
on demand production of a long string of entangled photons in a cluster
state \citep{Schwartz2016}. 

The main decoherence mechanism of the confined electronic spin (central
spin) in semiconductor QDs is its interaction with the spins of the
nuclei in its vicinity \citep{Gammon_2001,Efros2002,Khaetskii2002,Fischer2008}.
Therefore, it is essential, both scientifically and technologically,
to study and to characterize these dephasing processes. 

In this work we comprehensively study, both experimentally and theoretically,
the dephasing dynamics of QD confined electronic spins in 5 different
forms: a) conduction band electron, b) valence band heavy-hole, c)
negative trion, d) positive trion, and e) dark exciton (DE). All in
the same single QD.

Semiconductor QDs are formed by $\sim10^{5}$ molecules of one semiconductor
compound embedded in another semiconductor compound of higher bandgap
energy. These formations give rise to nanometer scale three-dimensional
(3D) potential traps, which confine single electronic charge carriers
(electrons in the conduction bands and holes in the valence bands)
and isolate them from their environment. The energy spectrum of these
confined carriers is therefore discrete, giving rise to well defined
and spectrally sharp optical transitions between these discrete levels
\citep{marzin1994,Dekel1998} . 

\begin{figure}
\includegraphics[width=1\columnwidth]{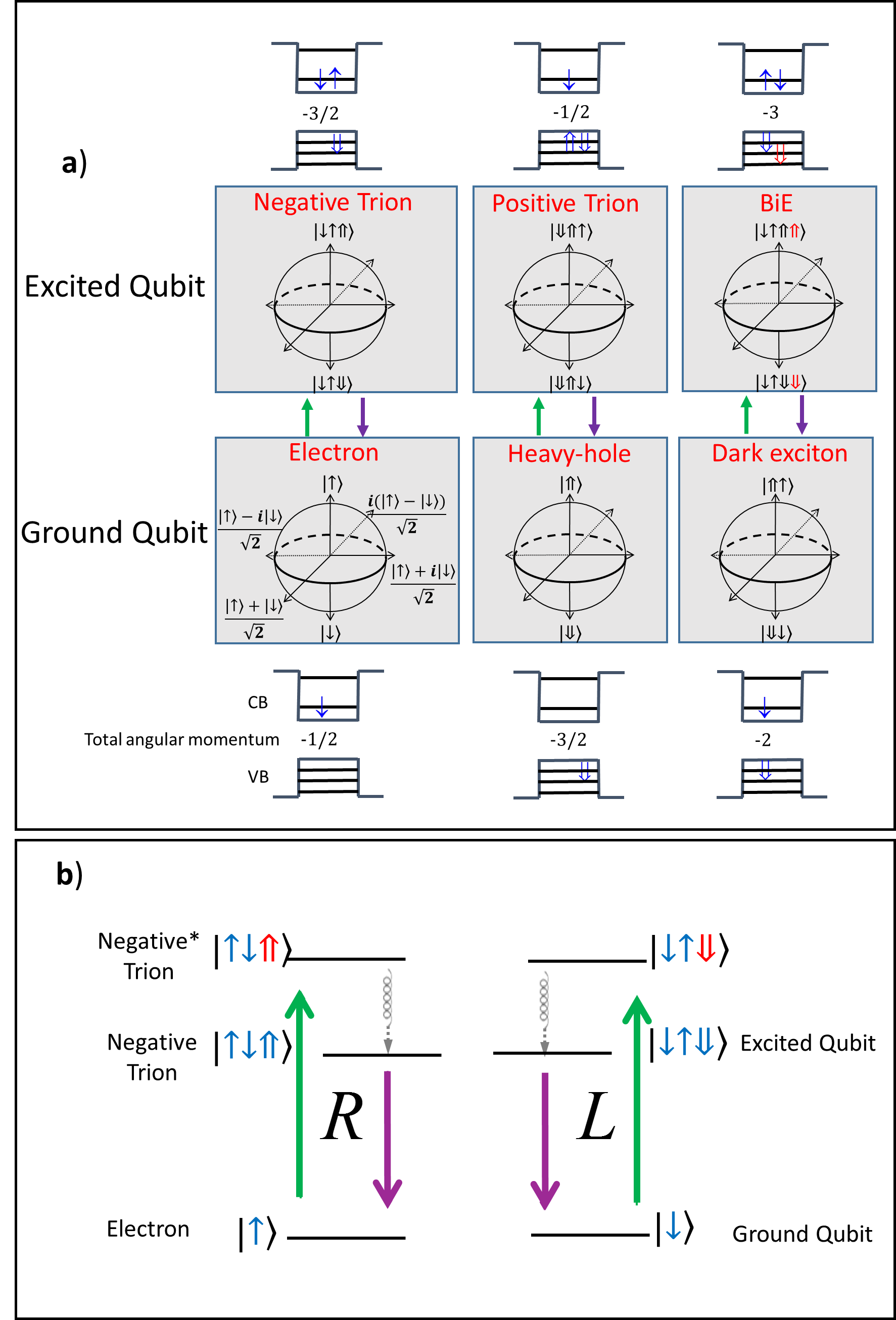}

\caption{\label{fig:Spin-wavefunctions}(a) Spin wavefunctions and Bloch-sphere
representations of the six matter spin qubits used in this work. The
6 qubits, represented by their Bloch spheres, are divided into 3 pairs
of ground and excited level qubits. The spin wavefunctions of the
ground- (excited-) level qubits are schematically described below
(above) the respective Bloch spheres, where $\uparrow$ ($\Downarrow$)
represents spin up electron (down heavy hole), and the blue (red)
color represents a carrier in its ground (excited) energy level. Green
upward arrows represent laser pulses which convert the ground level
qubit to its respective excited level qubit. Magenta downward arrows
represent single photons emitted from the excited qubits thereby returning
to the ground level qubit. (b) The optical transitions and polarization
selection rules for the electron-trion system, which form a ground
level–excited level qubit pair. Note that in this example (as in all
other cases) an optical $\Pi$-system is described, but the exciting
laser pulse is tuned to an excited trion level, in order to facilitate
polarization tomography of the emitted photon (magenta downdward arrows)
by spectrally separating the emission from the exciting laser pulse
(green upward arrows). The fast (\textasciitilde{}70 ps \citep{Schwartz2015})
phonon-assisted relaxation of the excited trion to the ground trion
level is represented by gray curly downward arrows. The right (left)
hand circular polarization of the photons which connect the $1/2$
($-1/2$) spin state of the ground level qubit with the $+3/2$ ($-3/2$)
spin state of the excited qubit are marked by $R$ ($L$). }
\end{figure}

In Fig.~\ref{fig:Spin-wavefunctions} we display the electronic spin
wavefunctions and Bloch-sphere representations of all the electronic
spin qubits used in this work. The confined conduction electron levels
have a vanishing atomic orbital momentum and thus their total spin
projection on the QD growth direction is $\pm1/2$ . Therefore, they
form physical two level systems or qubits \citep{DiVincenzo_2000}.
The spin state of the qubit is represented on the Bloch sphere, where
the spin up and the spin down states are located at the north and
south poles of the sphere, respectively, and any superposition of
these two states is represented by a point on the sphere's surface.
The confined valence-band electron states have total atomic orbital
momentum of 1. The spin-orbit interaction, together with the quantum
confinement along the growth direction and the biaxial lattice mismatch
compressive strain, inherent to our strain induced QDs, results in
a large energy splitting between the upper most valence states~\citep{Ivchenko2005}.
The highest valence electron states in which the orbital spin and
electronic spin are parallel, are few tens meV higher than the states
in which the orbital and electronic spins are anti-parallel. At low
temperature, the valence band states are fully occupied. Confined
positive charge carriers in the QD are therefore formed due to the
absence of valence band electrons. Thus, the lowest energy hole states
have angular momentum projection of $\pm3/2$ on the growth direction,
(heavy-holes). A heavy hole, is yet another form of a QD confined
electronic spin qubit~\citep{Brunner2009,De_Greve_2011} as shown
in Fig.~\ref{fig:Spin-wavefunctions}. Another form of a confined
electronic spin qubit is the electron-heavy-hole pair, or the exciton\citep{Benny2011,Kodriano2012}.
Excitons in which the heavy hole spin and the electron spin are anti-parallel
have total spin projection of $\pm1$, they are optically active and
therefore called bright excitons (BEs). The qubit that they form \citep{Benny2011,Poem2011,Kodriano2012}
recombines within a short radiative lifetime (\textasciitilde{}1 ns),
which limits their use as a matter spin qubit. In contrast, excitons
in which the electron and heavy-hole spins are parallel, are optically
inactive since the electromagnetic radiation barely interacts with
the electronic spin. These excitons are called dark excitons (DEs).
They have total spin projection of $\pm2$ on the QD growth axis and
live orders of magnitude longer than the BE \citep{McFarlane_2009}.
Consequently they can be used for implementing sophisticated quantum
information protocols \citep{Poem2010,Schwartz2015,Schwartz2016}. 

In the following we denote these three long lived forms of spin qubits
(electron, heavy-hole and DE) - ground level qubits. The ground level
qubits are stable, and once generated in the QD they live in it for
a very long time. The ground level qubits can be optically excited
to their respective excited level qubits by absorbing a single photon,
which adds an electron-hole pair to the QD. Moreover, by using a resonantly
tuned optical $\pi$-pulse, this excitation can be done deterministicaly.
The resonant excitation converts the ground level qubits to their
excited level qubits, as schematically described in Fig.~\ref{fig:Spin-wavefunctions}.
In Fig.~\ref{fig:Spin-wavefunctions}, green upward arrows represent
the optical laser excitations, which convert the electron spin qubit
to the negative trion qubit, the heavy-hole qubit to the positive
trion qubit, and the DE to the spin-blockaded biexciton (BiE)-qubit.
As can be seen in Fig.~\ref{fig:Spin-wavefunctions} the negative
and positive trion qubits, are formed by three carriers. The negative
trion is formed by two ground level conduction band electrons in a
singlet state and a single ground level heavy-hole, while the positive
trion is formed by two ground level heavy-holes and a single ground
level electron. In both cases, the spin state of the trion qubits
is determined by the minority carrier, $\pm3/2$ for the negative
trion, and $\pm1/2$ for the positive trion.

Unlike the trions , which are formed by three carriers, the BiE is
formed by four carriers. Two ground level electrons in a singlet spin
state, and two heavy holes with parallel spins in the ground and first
excited valence band levels. Consequently, the BiE qubit spin states
are $\pm3$, and it is determined by the two parallel heavy-holes'
spin directions. 

Once formed, the excited spin qubits, which are optically active,
decay radiatively within the radiative lifetime of a ground level
electron-hole pair (\textasciitilde{} 1 ns), by emitting a single
photon and the system returns to the ground level qubit. The photon
emissions are schematically described by the downward magenta arrows
in Fig.~\ref{fig:Spin-wavefunctions}.

If the upper qubit is properly initialized in a coherent superposition
of its two spin states, the polarization of the emitted photon (``flying
photonic qubit'') is expected to be entangled with the spin state
of the ground level spin qubit, which remains in the QD \citep{Gao2012,Greve2012,Schaibley2013,Schwartz2016}. 

At low temperatures and in the absence of external magnetic field,
the main decoherence mechanism of these electronic spin qubits is
the hyperfine interaction between the electronic (central) spin and
the spin of the nuclei of the $\sim10^{5}$ atoms which form the QDs
\citep{Gammon_2001,Efros2002,Khaetskii2002}. The two types of charge
carriers in semiconductors, the negative conduction band electrons,
and the positive, valence band holes interact differently with the
nuclei, since their orbital momentum around the nucleus is different.
The conduction electrons have zero atomic orbital momentum, while
valence band holes have unit atomic orbital momentum. Consequently,
the conduction electron's wavefunction strongly overlaps with the
nucleus and interacts with the nuclear spin via the Fermi contact
interaction. In contrast, the valence hole's wavefunction vanishes
at the nucleus site and therefore its spin interacts with the nuclear
spin via the weaker dipole-dipole hyperfine interaction \citep{Fischer2008}.
In addition, while the conduction-electron interaction with the nuclei,
which we denote by $\gamma_{e}$ is isotropic, the interaction of
the valence heavy hole for which the orbital angular momentum and
the spin are aligned parallel to the growth direction, is anisotropic.
We denote by $\gamma_{h_{z}}$the interaction of the valence heavy-hole
spin with the nuclei spin bath along the QD growth axis ($\hat{z}$)
and by $\gamma_{h_{p}}$ the interaction with nuclear spins in the
plane perpendicular to $\hat{z}$.

\begin{figure*}
\begin{centering}
\includegraphics[width=1\textwidth]{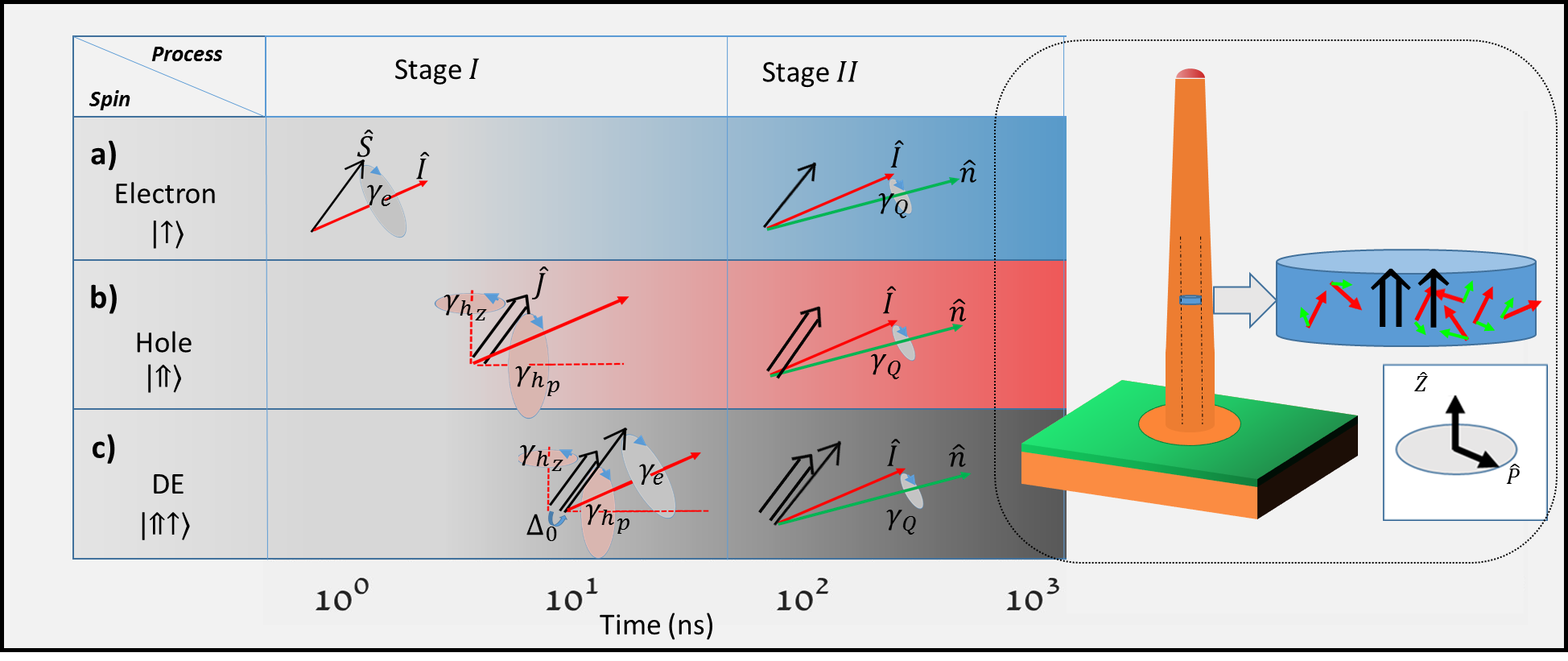}
\par\end{centering}
\caption{\label{fig:spin-dephasing}Schematic description of the spin dephasing
processes of the QD confined electron (a), heavy-hole (b), and dark
exciton (c). Each process is divided into two temporal stages: During
the first stage the initiated central spin precesses around the effective
magnetic field which results from the frozen fluctuation of the nuclear
spins of about $10^{5}$ atoms comprising the QD. The electron spin
($\hat{S)}$ interacts with the nuclear spin ($\hat{I}$) via the
isotropic Fermi contact interaction described by $\gamma_{e}$. The
heavy-hole spin ($\hat{J)}$ interacts with the nuclear spin via the
anisotropic dipole-dipole hyperfine interaction denoted by $\gamma_{h_{z}}$
and $\gamma_{h_{p}}$, where $\hat{z}$ is the QD growth direction
and $p$ denotes direction in a plane perpendicular to $\hat{z}$.
The dark exciton contains an electron and a heavy hole. Both spins
interact with the nuclear spin, and in addition, the electron and
hole interact with each other mainly via the isotropic exchange interaction
denoted by $\Delta_{0}$ \citep{Bayer2002}. During the second stage
the nuclear spins react to strain induced electric fields gradients
(EFG) in the QD \citep{Sinitsyn2012}. This interaction has a quadrupole
nature, and we denote it by $\gamma_{Q}$. The motion of each nuclear
spin is described by an effective local magnetic field in a direction
marked by $\hat{n}$ which the nuclear spin slowly precesses around.
During the second stage, we use the adiabatic approximation, by which
the central spin just follows the slowly varying effective nuclear
magnetic field. The various interaction magnitudes are summarized
and referenced in Table~\ref{tab:Interaction-energies}. Inset: schematic
description of the InAsP QD (blue) embedded in the InP photonic nanowire
(orange). The central spin is represented by the black arrows, the
nuclear spin bath are represented by the red arrows, and the EFGs
are schematically represented by green arrows in the magnified description
of the QD. }
\end{figure*}

The dynamics of the electronic central spin can be divided into two
different time domains as schematically described in Fig.~\ref{fig:spin-dephasing}
a, b and c for the electron, heavy hole and DE spins respectively
\citep{Efros2002}.

 During the first stage, the central spin precesses around a mean
effective magnetic field generated by the frozen fluctuations of the
nuclear spins in its vicinity. The electron interacts with the nuclear
field via the isotropic Fermi contact hyperfine interaction marked
by $\gamma_{e}$, while the heavy-hole interacts via the anisotropic
dipole-dipole hyperfine interaction marked by $\gamma_{h_{z}}$ and
$\gamma_{h_{p}}$. As the DE is formed by an electron-hole pair with
parallel spins, each of these carriers interacts with the nuclear
magnetic field, while at the same time they also interact with each
other, via the electron-hole exchange interactions. The most important
term in this interaction is the isotropic term $\Delta_{0}$ \citep{Bayer2002,Ivchenko2005},
separating the DE and BE (an antiparallel electron-hole pair) energy
levels. Being much stronger than the hyperfine interactions it prevents
the separate spin flip of either one of the two individual spins and
consequently protects the DE spin from dephasing. It turns out, as
we show in Appendix B, below, that the DE nuclear field induced dephasing
is caused mainly due to small DE-BE mixing terms (of order $10^{-3}$). 

During the second stage, at longer times, the fluctuations in the
nuclear magnetic field can no longer be considered ``frozen'' and
they slowly evolve in time. This evolution is described as local precession
of the effective magnetic field around local directions denoted by
$\hat{n}$. A relatively simple model describes this motion as generated
by the quadrupole interaction (denoted by $\gamma_{Q}$) of the nuclear
spins with the strain induced electric fields gradients in the QD
\citep{Sinitsyn2012,Bechtold2015}. We adopt this description, since
it permits analytic solution to the problem, thereby simplifying the
comparison with the measured data, while keeping the generality of
our approach. Finally, at yet longer times, which is beyond the scope
of this work, the nuclei also interact with each other via the dipole-dipole
nuclear interaction~\citep{Erlingsson2004}. During the second stage
the central spin continues to interact with the slowly varying effective
nuclear magnetic field in the same manner as it does during the first
stage. Therefore, the central spin dynamics can be described as a
sort of ``convolution'' between the relatively fast dynamics of
the spin around the average nuclear magnetic field, with the dynamics
of the slowly varying nuclear field.  

The details of the model involved in these calculations, which follows
references \citep{Efros2002,Sinitsyn2012,Bechtold2015}, describing
the evolution of the electron, and the generalization of the model
to include the heavy-hole evolution, are described in Appendix~A.
The model which describe the dynamics of the DE is developed in Appendix~B. 

A great deal of effort was devoted to study the coherence properties
of the central electronic spin for both, conduction band electrons
\citep{Bluhm2010,Braun2005}, and valence band heavy-holes \citep{Brunner2009,Eble2009,Fras2011,Li2012,Gerardot2008},
confined in QDs. The temporal evolution of a single electron spin
at vanishing external magnetic field was experimentally measured recently
by Bechtold and coworkers \citep{Bechtold2015}. To the best of our
knowledge, similar measurements for the heavy-hole as a central spin
have not been reported so far. Here, we present comprehensive measurements
of the spin depolarization dynamics for both the electron and the
heavy hole as well as for their correlated pair – the DE. All these
forms of central electronic spin are confined to the same QD. In addition,
we show, by measuring the temporal evolution of the positive and negative
trions' spins, that the presence of two additional paired charge carriers,
does not affect the central spin depolarization. Our measurements
were preformed optically without applying any external magnetic field.
In addition, we carried out the experiments in a way which prevented
the generation of a steady state nuclear Overhauser field. The experimental
methods and measurements are described below and the measured results
are compared with the central spin models discussed in the Appendices. 

\section{The device and experimental methods}

The InP nanowire containing a single InAsP quantum dot~\citep{Dalacu2009,Dalacu2012,Bulgarini2014}
was grown using chemical beam epitaxy with trimethylindium and pre-cracked
$\mathrm{PH_{3}}$ and $\mathrm{AsH_{3}}$ sources. The nanowires
were grown on a $\mathrm{SiO_{2}}$-patterned (111)B InP substrate
consisting of circular holes opened up in the oxide mask using electron-beam
lithography and a hydrofluoric acid wet-etch. Gold was deposited in
these holes using a self-aligned lift-off process, which allows the
nanowires to be positioned at known locations on the substrate. The
thickness of the deposited gold is chosen to give 20-nm to 40-nm diameter
particles, depending on the size of the hole opening. The nanowires
were grown at $420^{\circ}$ C with a trimethylindium flux equivalent
to that used for a planar InP growth rate of 0.1 $\text{μm/hr}$
on (001) InP substrates at a temperature of $500^{\circ}$ C. The
growth is a two-step process: (i) growth of a nanowire core containing
the quantum dot, nominally 200 nm from the nanowire base, and (ii)
cladding of the core to realize nanowire diameters (around 200 nm)
for efficient light extraction. The quantum dot diameters are determined
by the size of the nanowire core. The particular QD reported on here
has diameter of $\sim30$ nm.

The sample was placed inside a sealed metal tube cooled by a closed-cycle
helium refrigerator maintaining a temperature of 4 K. A \texttimes 60
microscope objective with numerical aperture of 0.85 was placed above
the sample and used to focus the laser beams on the sample surface
and to collect the emitted PL from it. Pulsed laser excitations were
used. The picosecond pulses were generated by two synchronously pumped
dye lasers at a repetition rate of 76 MHz. The temporal width of the
pulses was 12~ps and their spectral width $\sim100$~μeV. Light
from a continuous wave (CW) laser, modulated by an acousto-optic modulator,
synchronized with the dye lasers, was used to produce pulses of up
to 30 ns duration. These pulses were used to set the average QD charge
state \citep{Benny2012}. A second CW laser, modulated by an electro-optic
modulator, was used to produce depletion pulses of 30 ns duration
\citep{Schmidgall2015}. The timing between the two synchronized ps
pulses was controlled using 2 cavity dumpers which effectively reduced
the repetition rate down to 0.5~MHz. In addition, a computer controlled
motorized delay line was used to finely tune the temporal delay between
the pulses. The polarizations of the excitation pulses were independently
adjusted using polarized beam splitters (PBS) and two pairs of computer-controlled
liquid crystal variable retarders (LCVRs)~\citep{Schwartz2016}.
The collected PL was equally divided into 2 beams by a non-polarizing
beam splitter. Two pairs of LCVRs and a PBS were then used to analyze
the polarizations of each beam. This way the emitted PL was divided
into four beams, allowing selection of two independent polarization
projections and their complementary polarizations. The PL from each
beam was spectrally analyzed by either a 1 or 0.5 meter monochromator
and detected by a silicon avalanche photodetector coupled to a PicoQuant
HydraHarp 400™ time-correlated photon counting and time tagging system,
synchronized with the pulsed lasers. This way the arrival times of
up to 4 emitted photons have been recorded with respect to the synchronized
laser pulses.

\begin{figure}
\begin{centering}
\includegraphics[width=1\columnwidth]{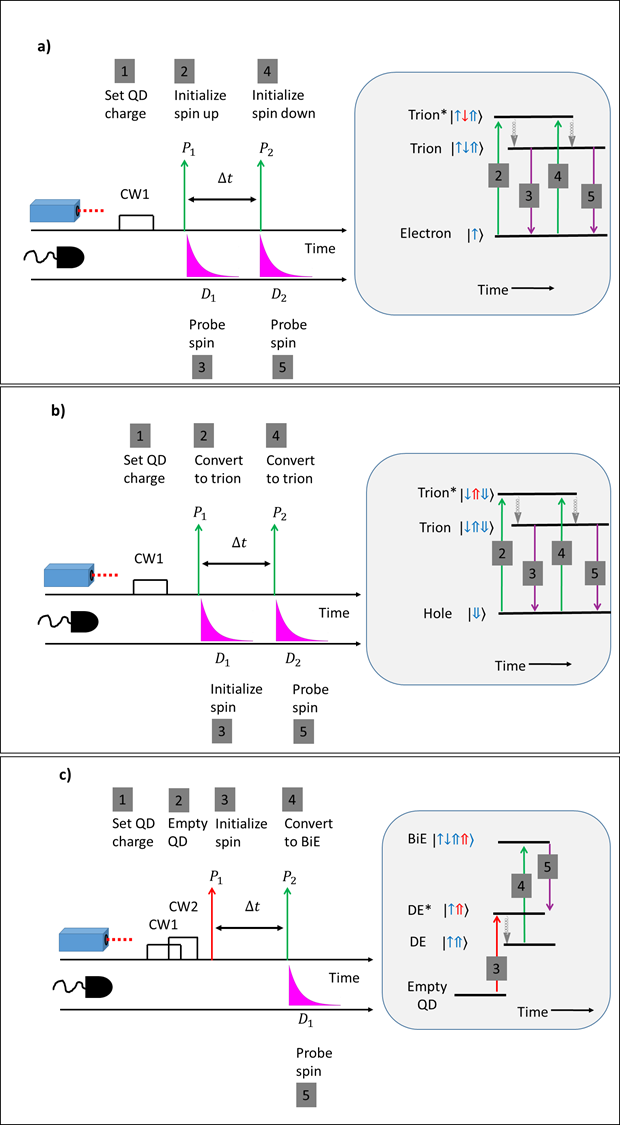}
\par\end{centering}
\caption{\label{fig:Experimental-procedures}Schematic description of the experiments
for measuring the spin dynamics of: a) Positive and negative trions,
b) Single electron and single heavy-hole. c) Dark exciton. The optical
transitions in each experiment are described by the energy level diagram
to the right. The carrier's spins are marked in the figure using the
notations of Fig.~\ref{fig:Spin-wavefunctions}, where blue (red)
color represent ground (excited) single carrier states. CW1 and CW2
represent the 20~ns gated CW laser pulses where $P_{1}$ and $P_{2}$
represent the 12~ps pulses produced by the synchronously pumped and
cavity dumped dye lasers. $\Delta t$ is the time delay between the
two pulses in each repetition period, controlled by two cavity dumpers
and a delay line. $D_{1}$ and $D_{2}$ represent the emission and
time resolved detection of the two single photons emitted as a result
of the $P_{1}$ and $P_{2}$ excitations. }
\end{figure}

We used the optical transitions between the ground level qubits and
the excited level qubits to initialize the spin state of both qubits,
and then for probing the spin state of the qubits at a later time.
We facilitate the optical transition selection rules of the $\Pi$-systems
described in Fig.~1b in order to do that.

For initializing the excited qubit, one simply applies an $R$ or
$L$ polarized $\pi$-pulse. For probing the excited qubit spin projection,
one simply measures the degree of circular polarization of the emitted
photons $\hat{S}_{z}=\left(I_{R}-I_{L}\right)/\left(I_{R}+I_{L}\right)$
where $I_{R(L)}$ is the measured emission intensity projected on
right (left) hand circular polarization.

The initialization of the ground level qubit is provided by detecting
$R$ or $L$ polarized single photon, which heralds the spin state
of the qubit at the photon emission time. Probing the ground level
qubit spin state is done by first converting the state into the state
of the excited level qubit, using an horizontally linearly polarized
($H=\left(R+L\right)/\sqrt{2}$) $\pi$-pulse, and then measuring
the time resolved degree of circular polarization of the emitted photons.
For example, in Fig.~\ref{fig:Spin-wavefunctions}(b) if the electron
spin state before the pulse is described by: $\hat{\rho}_{\mathrm{electron}}=p\ket{\Psi_{\mathrm{electron}}}\bra{\Psi_{\mathrm{electron}}}$+$\left(1-p\right)\frac{1}{2}\mathbb{I}$,
where $\mathbb{I}$ is the identity matrix and $p$ is the probability
of $\hat{\rho}_{\mathrm{electron}}$ being in a pure state $\ket{\Psi_{\mathrm{electron}}}=\alpha\ket{\uparrow}+\beta\ket{\downarrow}$,
then after the pulse the photogenerated trion spin state is given
by: $\hat{\rho}_{\mathrm{trion}}=p\ket{\Psi_{\mathrm{trion}}}\bra{\Psi_{\mathrm{trion}}}$+$\left(1-p\right)\frac{1}{2}\mathbb{I}$,
with $\ket{\Psi_{\mathrm{trion}}}=\alpha\ket{\uparrow\downarrow\Uparrow}+\beta\ket{\downarrow\uparrow\Downarrow}$,
with the same $\alpha$, $\beta$ and $p$. Here, we assume of course,
that the fidelity of the optical excitation by the $H$ polarized
$\pi$-pulse is unity and that the experimental deviation from truly
$H$ polarization is negligible. The spin projection of excited qubit
on the $\hat{z}$-direction is then deduced by measuring the degree
of circular polarization of the emitted photons. 

We conducted 5 different experiments in order to comprehensively study
the central spin dynamics for various confined spin qubits in the
QD. In the first 2 measurements, schematically described in Fig.~\ref{fig:Experimental-procedures}a,
we measured the depolarization of the negative or positive trions.
We first pump the QD to either a negative or a positive charge state
by using above bandgap CW1 pulse of about 10ns duration\citep{Benny2012}.
Then, either an excited negative or positive trion was photogenerated
by using a short circularly polarized quasi-resonant \textasciitilde{}12
ps long laser pulse. The polarization of the excitation pulse determines
the spin polarization of the minority carrier in the initialized trion
{[}hole (electron) in the negative (positive) trion{]}. After a fast
(\textasciitilde{}70 ps \citep{Schwartz2015}) spin preserving phonon
assisted relaxation of the excited trion, a ground level trion is
formed. When the trion decays radiatively, the polarization of the
emitted photon reflects the spin of the minority carrier at the particular
time in which the photon is emitted. Thereby, by using time resolved
circular polarization sensitive PL measurements we probe the spin
relaxation dynamics of the minority carrier in the trion. This technique
provides a simple way of measuring the dynamics of the spin of the
confined electron (hole) in the presence of a spin singlet pair of
two holes (electrons). Unfortunately, this simple method is limited
by the relatively short radiative lifetime of the trion. Only the
evolution during the first time domain can be measured this way. In
order to avoid generating a steady state Overhauser field in the QD
due to the repeated circularly polarized quasi-resonant excitation
pulse, a second pulse with opposite circular polarization is used
to re-excite the trion a few ns after the first pulse, during the
same excitation period. The time resolved degree of circular polarization
was deduced using the resulted PL from both complementary pulses. 

The measurement of the spin dynamics of either the single electron
or heavy-hole was carried out using the same experimental system but
at somewhat different manner, as schematically described in Fig.~\ref{fig:Experimental-procedures}b.
In the inset to this figure we describe the energy levels of the heavy-hole
system. Here, after the optical charging, a trion was generated by
quasi resonant excitation using a horizontal ($H$) polarized pulse.
Either the electron or the hole spin was initialized by detecting
the circular polarization of the emitted single photon. In order to
probe the temporal dependence of the spin state of the carrier, a
second, horizontal polarized delayed 12~ps pulse is used to re-excite
the carrier to its respective trion and the resulting circular polarization
of the emitted photon is used to measure the spin polarization of
the carrier at the re-excitation time. This measurement is not limited
by the radiative lifetime of the trion, however, it requires two-photon
intensity correlation measurements in a relatively slow repetition
rate (\textasciitilde{}500~kHz). We achieved this low repetition
rate by using the cavity dumpers. The feasible maximal delay time
(\textasciitilde{}1~μs$)$ between the pulses was defined by the
rejection ratio (of about $\sim2\vartimes10^{-3}$ ) of neighboring
pulses of the cavity dumpers. Note that in these experiments the generation
of an Overhauser field is avoided because the initialization of the
central spin is not done deterministically by using circularly polarized
excitation, but rather probabilistically by post-selecting the detected
circular polarization of the emitted first photon. 

The spin dynamics of the DE was probed as schematically described
in Fig.~\ref{fig:Experimental-procedures}c. Here, we used above-bandgap
optical pumping of about 20~ns to neutralize the QD and then another
quasi-resonant pumping of about 20~ns to deplete the QD from the
DE \citep{Schmidgall2015}. After depleting the QD, a quasi-resonant
circularly polarized 12~ps pulse initialized the DE in spin up excited
state \citep{Schwartz2015a}. Following this initialization, the DE
relaxes to its ground state within \textasciitilde{}70 ps by spin-preserving
emission of a phonon. In order to probe the DE state, a delayed, linearly
polarized resonant 12 ps pulse converted the DE qubit into the BiE
qubit. Note that the horizontal polarization of the laser preserves
the phase of the qubit. The detection of a circularly polarized photon,
which results from the radiative recombination (\textasciitilde{}1
ns lifetime) of the BiE is then used to probe the spin state of the
DE in the QD, at the converting pulse time. Repetition rates as low
as \textasciitilde{}500~kHz, allow temporal delays of over 1~μs
between initialization and probing of the spin. In this experimental
method an Overhauser field is not generated in the sample since the
gated CW pulses used to optically pump and deplete the QD are linearly
polarized. 

\begin{figure*}
\begin{centering}
\includegraphics[width=1\textwidth]{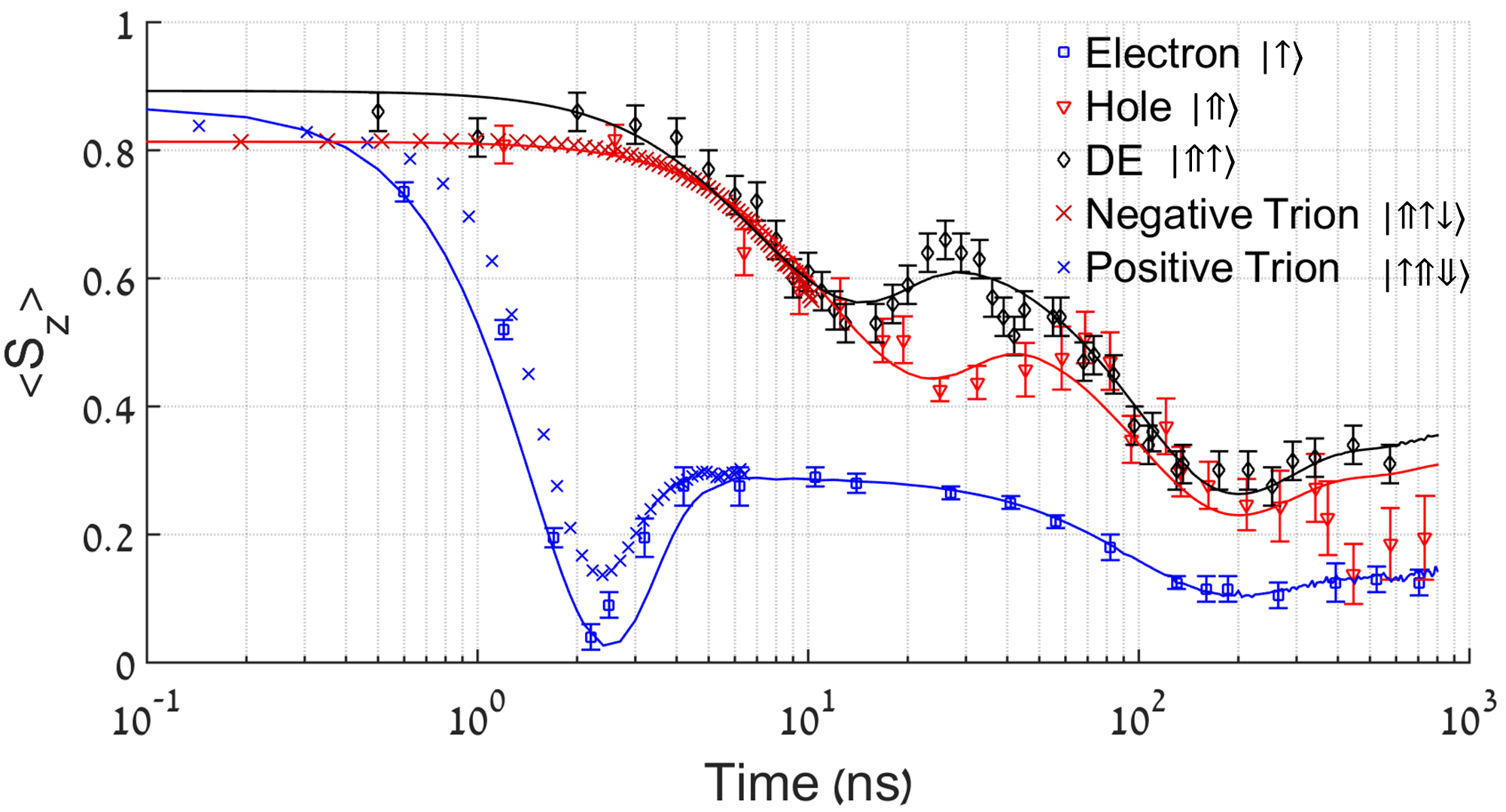}
\par\end{centering}
\caption{\label{fig:Results}Measured central spin polarization $\left\langle S_{z}\right\rangle $
as a function of time after its initialization, for the QD confined
electron (blue square symbols), heavy-hole (red triangle symbols),
dark exciton (black diamond symbols), negative trion (red $\times$symbols),
and positive trion (blue $\times$ symbols). Error bars represent
one standard deviation of the experimental uncertainty. Solid color
matched lines represent the fitted theoretical model (see appendices),
for each case. The spin wavefunctions are schematically described
in the legend, where the notations of Fig.~\ref{fig:Spin-wavefunctions}
are used. }
\end{figure*}

\section{Results and discussion}

In Fig.~\ref{fig:Results} we present the measured degree of the
average central spin polarization $\left\langle S_{z}(t)\right\rangle $
as a function of time after its initialization, for the 5 spin qubits:
the conduction band electron, the valence band heavy-hole, the positive
and negative trions, and the dark exciton. The error bars represent
one standard deviation of the experimental uncertainty. At time zero
the central spin is initialized to spin-up state. Then, the projection
of the spin on $\hat{z}$ direction (the QD growth axis) is displayed
as a function of time.

The conduction band electron spin state (blue rectangles) depolarizes
from its initial state within \textasciitilde{}2~ns. The spin polarization
then revives to about a third of the initial polarization. From this
level the polarization continues to decay at a much slower rate, reaching
a second minimum at about \textasciitilde{}200~ns. Afterwards the
spin polarization revives again to about 10\% of the initial polarization.
This behavior is similar to that reported in Ref.~\citep{Bechtold2015},
as predicted by Ref.~\citep{Efros2002}. Roughly speaking, the first
fast dephasing step is a measure for the strong Fermi-contact hyperfine
interaction of the electron with the nuclear spin bath, while the
second step measures the strength of the quadrupole interaction of
the nuclear spin bath with the strain induced electric field gradients
in the QD. 

After initialization, the heavy-hole (red triangles) spin depolarizes
in about an order of magnitude slower than the electron spin. This
is due to the much weaker dipole-dipole hyperfine interaction. The
hole spin polarization decreases at about \textasciitilde{}20 ns to
about one half of its initial polarization. Afterwards it mildly revives
followed by a slow decay due to the quadrupole interaction of the
nuclear bath. 

The positive trion spin polarization (blue $\times$ symbols), behaves
similarly to that of the electron, while the negatively charged trion
spin polarization (red $\times$ symbols) follows that of the heavy-hole.
This is not surprising, since the trion polarization reflects the
polarization of the unpaired minority carrier, in the presence of
the two paired majority carriers. As explained above, the trions spin
measurements are limited by their radiative lifetime of about \textasciitilde{}1~ns.

The dark exciton (black diamonds) decoheres slowly, in a similar rate
to the heavy-hole. However, like the electron, after the initial decay,
it strongly revives to about two thirds of its initial polarization.
This is due to the strong exchange interaction between the electron
and hole that protects both carriers from flipping their individual
spins. Later, after \textasciitilde{}200~ns the dark exciton polarization
continues to decay due to the quadrupole interaction.

We fit the measured temporal behavior of the electron, heavy hole
and dark exciton using one conceptually simple central spin model.
For the fitting, only five free parameters are used: 1) The hyperfine
Fermi-contact interaction $\gamma_{e}$, 2) the heavy-hole out-of-plain
hyperfine dipole-dipole interaction $\gamma_{h_{Z}}$, 3) the heavy-hole
in-plain hyperfine dipole-dipole interaction $\gamma_{h_{p}}$, 4)
the DE in-plane interaction $\gamma_{\mathrm{DE}_{p}}$ and 5) the
quadrupole interaction $\gamma_{Q}$. These parameters are accurately
defined in the appendices, where the models are discussed for the
electron and the heavy-hole (Appendix A), and for the DE (Appendix
B). 

The best fitted parameters are given in Table I, where they are also
compared with the available literature. Our analysis provides an estimation
of the number of atoms in the QD: $N_{L}=3\cdot10^{5}.$ With this
estimation our fitted hyperfine Fermi contact $\gamma_{e}$ is comparable
to that of Ref.~\citep{Gotschy1989}.

Characteristic spin depolarization times during the first and second
temporal stages can be obtained from our fitting procedure quite straightforwardly.
Since the central spins in this work are initialized in z direction,
depolarization is caused by the in-plane interaction parameters. Thus,
the temporal location of the first minimum is a rough measure of the
in-plane interaction parameter: $T_{\min}=\hbar/\gamma_{_{p}}\sim2$,
20 and 14~ns for the electron, heavy-hole and DE, respectively. Thus,
$\gamma_{e}$,~$\gamma_{h_{p}}$ and~$\gamma_{\mathrm{DE}_{p}}$
are given by $0.34,$ $0.031$ and $0.047\,\text{μeV}$, respectively. 

The central spin interaction with the nuclear field along the $z$-direction,
acts as a restraining force, which actually prolongs the spin coherence.
Therefore, roughly speaking, the ratio between these interactions
( $R_{\gamma}=\gamma_{_{Z}}/\gamma_{_{P}})$ determines the depth
of the first polarization minimum and the maximum value of the polarization
after its revival. We thus obtain $R_{\gamma}$=1, 3.5 and 5, for
the electron, hole and DE, respectively. Note that for the electron
the ratio is by definition 1, and therefore the polarization degree
revives to 1/3 of its initial value, while for the hole and DE it
revives to higher values. During the second temporal stage, the polarization
of all three central spins decays more or less at the same rate, determined
by the quadrupole interaction $\gamma_{Q}$. Therefore the temporal
location of the second minimum is about the same in all cases given
by $T_{\min_{Q}}=\hbar/\gamma_{Q}\sim200\,\mathrm{ns}$ or $\gamma_{Q}\sim0.003\,\text{μeV}$.

A common practice for quantifying the depolarization value of a spin
qubit is to define the depolarization time as the time it takes for
the polarization to reduce to 1/e of its initial state. We adopt this
practice, though the measured depolarizations are clearly non-exponential.
The measured depolarization times thus obtained are $1.5$, $130$
and $145$~ns for the electron, heavy hole and DE respectively.

\begin{table}
\centering{}\caption{\label{tab:Interaction-energies}The various interaction energies
as obtained by the best model fitting to the measured temporal depolarization
of five different electronic spin qubits confined in the same QD.
The theoretical model and the fitted parameters are described in the
text and the Appendices.}
\medskip{}
\begin{tabular}{llc}
\toprule 
Interaction~~~~ & This work (μeV)~~~~~ & Literature (μeV)~~~~~\tabularnewline
\midrule
\addlinespace
$\gamma_{e}\,$ & $0.34\pm0.03$ & $0.33$ \citep{Bechtold2015}\tabularnewline\addlinespace
\addlinespace
\midrule 
$\gamma_{h_{z}}\,\,$ & $0.11\pm0.03$ & $0.081$ \citep{Eble2009}\tabularnewline
\addlinespace
$\gamma_{h_{p}}\,\,$ & $0.031\pm0.006$ & $0.047$ \citep{Eble2009}\tabularnewline\addlinespace
\addlinespace
$\gamma_{\mathrm{DE}_{p}}\,\,\,$ & $0.047\pm0.006$ & —–\tabularnewline\addlinespace
\addlinespace
\midrule 
$\gamma_{Q}\qquad$  & $0.0031\pm0.001$  & $0.00087$ \citep{Bechtold2015}\tabularnewline
\bottomrule
\addlinespace
\end{tabular}
\end{table}

\section{Summary}

We investigated both experimentally and theoretically the depolarization
dynamics of five different electronic spin configurations  confined
in the same semiconductor quantum dot. Our measurements were carried
out all optically and in the absence of externally applied magnetic
field. We show that the measured temporal spin depolarization is well
described by a central spin model which attributes the depolarization
to the hyperfine interaction between the electronic spin and the nuclear
spin bath of the QD atoms. 

We divide the depolarization into two temporal stages. During the
initial stage the central spin precesses around the effective magnetic
fields of the frozen fluctuations of the $10^{5}$ nuclear spins in
the QD. During the second stage the central spin precession follows
adiabatically the nuclear spin bath dynamics which ceases to be frozen
and effectively precesses around strain induced electric fields gradients
in the QD.

These two processes result in a relatively fast initial depolarization
of the central spin reaching a first minimum. The depolarization minimum
is then followed by a temporal revival of the polarization degree
and finally by a second depolarization reaching a minimum at a much
later time which is more or less equal for all the electronic central
spin cases. 

Our model assumes that while the hyperfine interaction between the
central spin and the nuclear spins is isotropic for the electron,
it is anisotropic for the heavy-hole and therefore also for the DE,
which is formed by an electron–heavy-hole pair. The depolarization
times that we measured in zero magnetic field show that the electron
depolarizes much faster than the heavy-hole This observation is explained
by the difference between the strong isotropic electron-nucleous hyperfine
contact interaction ($\gamma_{e}$ ) and the anisotropic hole-nucleous
dipolar hyperfine interactions ($\gamma_{h_{Z}},\gamma_{h_{p}}$).
The heavy hole spin depolarizes faster than the dark exciton spin
due to the electron-hole exchange interaction, which protects the
dark-exciton spin from depolarizing. The depolarization of the dark-exciton
results from residual dark exciton–bright exciton mixing. We believe
that this mixing can be significantly reduced by increasing the QD
symmetry and by avoiding alloying. In this case the dark-exciton may
form an almost non-dephasing electronic spin qubit in a semiconductor
environment.

\section{Acknowledgement}

The support of the Israeli Science Foundation (ISF), and that of the
European Research Council (ERC) under the European Union’s Horizon
2020 research and innovation programme (grant agreement No 695188)
are gratefully acknowledged. 


\appendix

\section{Hyperfine interaction of the electron and the heavy-hole}

We outline here a model for describing the temporal evolution of the
QD confined central spin polarization in the absence of externally
applied magnetic field but in the presence of effective magnetic field
generated by the nuclear spins, which comprise the QD. As the central
spin we consider either the electron or the heavy hole. We then apply
the same model also to a central spin formed by the DE – a long lived
electron–heavy-hole pair, as will be discussed in Appendix B. 

As all three cases involve a two level system (a qubit) they may be
described using the Pauli matrices $\sigma_{x},\sigma_{y},\sigma_{z}$
and the effective Hamiltonian must take the form 
\[
H=\frac{1}{2}\vec{C}\cdot\vec{\sigma},
\]
for some $\vec{C}=(C_{x},C_{y},C_{z})$. The exact expression of $\vec{C}$
will be different, of course, for each type of central spin. 

The hyperfine Fermi-contact interaction between an electron and all
the nuclei in the QD is given by \citep{Efros2002} :
\[
H=\frac{\nu_{0}}{2}\sum|\psi_{\mathrm{env}}(\vec{r}_{i})|^{2}A_{e}^{i}\vec{I}_{i}\cdot\overrightarrow{\sigma}.
\]
Here $\nu_{0}$ is the volume of the unit cell, $\vec{r}_{i}$ and
$\vec{I}_{i}$ are the ith nucleus position and its spin operator,
$\psi_{\mathrm{env}}(\vec{r})$ describes the electron envelope wavefunction
and $A_{e}^{i}$ is an effective hyperfine interaction constant between
the electron and the specific nucleus in the $\vec{r}_{i}$ position
where the index i runs over all the nuclei in the QD. Since $A_{e}^{i}$
depends on the atomic nuclear spin it is much larger for indium atoms
than for all other atoms in the QD. Thus, in principle, one can neglect
other nuclei contributions. We proceed by defining an expression for
the effective magnetic field, which the nuclei apply on the electron.
The field, known also as the Overhauser field, is defined as: 
\[
\vec{B}_{N}=\frac{1}{g_{e}\mu_{B}}\vec{C}_{e}=\frac{\nu_{0}}{g_{e}\mu_{B}}\sum A_{e}^{i}|\psi_{\mathrm{env}}(\vec{r}_{i})|^{2}\left\langle \vec{I}_{i}\right\rangle _{N},
\]
where $g_{e}$ and $\mu_{B}$ are the electron g-factor and Bohr magneton
respectively, and $\left\langle ...\right\rangle _{N}$ denotes a
quantum mechanical average over the nuclear spins which interact with
the electron.

Assuming that different nuclear spins are not correlated allows one
to treat $\vec{B}_{N}(t)$ as having isotropic Gaussian random distribution
satisfying 
\[
\langle\vec{B}\rangle=0,\quad\langle B_{Nx}^{2}\rangle=\langle B_{Ny}^{2}\rangle=\langle B_{Nz}^{2}\rangle=\sigma^{2},
\]
where the width of the distribution $\sigma$ is given by \citep{Efros2002}
\[
3\sigma^{2}=\sum\frac{(A_{e}^{i})^{2}}{\mu_{B}^{2}g_{e}^{2}}\nu_{0}^{2}|\psi_{\mathrm{env}}(\vec{r}_{i})|^{4}I_{i}(I_{i}+1).
\]
 It is then convenient to define a modified unitless magnetic field
$\vec{\tilde{B}}=\frac{1}{\sigma}\vec{B}_{N}$. In the following we
simply mark this modified Overhauser field as $\vec{B}$. The electron
spin Hamiltonian can then be expressed by $H=\frac{1}{2}\vec{C}_{e}\cdot\vec{\sigma}$
with $\vec{C}_{e}=\gamma_{e}\vec{B}$ where $\gamma_{e}=g_{e}\mu_{B}\sigma$
is the electron coupling constant in energy units, which we use as
a fitting parameter.

While for the electron, $s$-wave molecular symmetry results in a
scalar effective coupling $A_{e}^{i}$, for the heavy hole it is described
by an anisotropic tensor 
\[
\hat{A}_{h}^{i}=\left(\begin{array}{ccc}
A_{h,p}^{i}\\
 & A_{h,p}^{i}\\
 &  & A_{h,z}^{i}
\end{array}\right).
\]
Where the in plane dipole-dipole interaction constant $A_{h,p}^{i}$
does not strictly vanish for the heavy-hole due to mixing with the
light-hole \citep{Eble2009}. Therefore, for the heavy-hole we define
$C_{z}=\gamma_{h_{z}}B_{z},C_{x,y}=\gamma_{h_{p}}B_{x,y}$ where $\gamma_{h_{z}}>\gamma_{h_{p}}$,
are also fitting parameters. Strictly speaking, the field $\vec{B}$
appearing here is not exactly the same one as in the electron case.
This is due to differences in relative weighting of various nuclei
between electron and hole wavefunctions. For our purpose, however,
it is sufficient that the fields have the same Gaussian statistics.
For the moment we allow the functional relation between $\vec{C}$
and $\vec{B}$ to be arbitrary and since our discussion is independent
of these relations, it applies to all three cases.

At short times $\vec{B}$ and hence also $\vec{C}$ can be treated
as time independent and one readily find the solution 
\begin{align}
\vec{S}(t) & =\frac{\vec{S}_{0}\cdot\vec{C}}{C^{2}}\vec{C}+\left(\vec{S}_{0}-\frac{\vec{S}_{0}\cdot\vec{C}}{C^{2}}\vec{C}\right)\cos\left(\frac{C}{\hbar}t\right)\label{st}\\
 & -\frac{\vec{S}_{0}\times\vec{C}}{C}\sin\left(\frac{C}{\hbar}t\right),\nonumber 
\end{align}
where $\vec{S}_{0}=\vec{S}(0)$ is the central spin initial value.
The first term is time independent and survives for long times. Upon
averaging over the random ensemble of possible $\vec{C}$s one typically
finds that the oscillating terms turn into exponentially decaying
transients, relevant at short times only. In practice the last term
usually vanishes by symmetry under $\vec{C}\rightarrow-\vec{C}$.
In particular it applies to our experiments, which were carried out
in the absence of externally applied magnetic field. Therefore, in
the following we disregard this term.

At longer times, we use the adiabatic approximation and assume that
the central spin follows the direction of $\vec{C}$, while the rapidly
rotating components orthogonal to $\vec{C}$ average to zero. We can
therefore write 
\begin{equation}
\vec{S}(t)=\left(\vec{S}_{0}\cdot\hat{C}(0)\right)\hat{C}(t)=\frac{\vec{S}_{0}\cdot\vec{C}(0)}{C(0)C(t)}\vec{C}(t).\label{lt}
\end{equation}
For small $t$ this clearly coincides with the first term of Eq.~(\ref{st}).
As the other terms of Eq.~(\ref{st}) vanish at long times one sees
that the two relations Eqs.~(\ref{st},\ref{lt}) can be combined
into an expression which applies at arbitrary time $t$: 
\begin{align}
\vec{S}(t) & =\frac{\vec{S}_{0}\cdot\vec{C}(0)}{C(0)C(t)}\vec{C}(t)\\
 & +\left(\vec{S}_{0}-\frac{\vec{S}_{0}\cdot\vec{C}(0)}{C(0)^{2}}\vec{C}(0)\right)\cos\left(\frac{C(0)}{\hbar}t\right).\nonumber 
\end{align}

The Gaussian probability density corresponding to the dimensionless
Overhauser field at a given moment is given by 
\begin{equation}
dP_{1}=\frac{1}{(2\pi)^{3/2}}\exp\left(-\frac{1}{2}B^{2}\right)d^{3}B.\label{p1}
\end{equation}
Assuming further that 
\[
\langle B_{i}(t_{1})B_{j}(t_{2})\rangle=\delta_{ij}f(t_{2}-t_{1})
\]
(Consistency requires $f(0)=1$) we can write the joint probability
density of $\vec{B}_{1}=\vec{B}(0)$ and $\vec{B}_{2}=\vec{B}(t)$
as

\begin{align}
dP_{2} & =\frac{d^{3}B_{1}d^{3}B_{2}}{\left(2\pi\sqrt{1-f(t)^{2}}\right)^{3}}\nonumber \\
 & \exp\left[-\frac{1}{2}\left(B_{1}^{2}+B_{2}^{2}-2f(t)\vec{B}_{1}\cdot\vec{B}_{2}\right)/(1-f(t)^{2})\right].\label{eq:dp2}
\end{align}

Using the probability distributions Eqs.~(\ref{p1},\ref{eq:dp2})
we can write the average central spin evolution as

\begin{align}
\langle\vec{S}(t)\rangle & =\int\frac{\vec{S}_{0}\cdot\vec{C}(0)}{C(0)C(t)}\vec{C}(t)\,dP_{2}\nonumber \\
 & +\int\left(\vec{S}_{0}-\frac{\vec{S}_{0}\cdot\vec{C}}{C^{2}}\vec{C}\right)\cos\left(\frac{C}{\hbar}t\right)dP_{1}\label{eq:St}
\end{align}
Actual computation of the integrals requires using the specific functional
relation between $\vec{B}$ and $\vec{C}$.

For the electron as the central spin, we simply substitute $\vec{C}=\gamma_{e}\vec{B}$
and $\vec{S}_{0}=\hat{z}$ in Eq.~(\ref{eq:St}) and obtain integrals
which can be evaluated analytically \citep{Efros2002,Bechtold2015},
resulting in
\begin{align*}
\langle S_{z}\rangle & =\frac{2}{3}\left(1-\left(\frac{\gamma_{e}t}{\hbar}\right)^{2}\right)e^{-\frac{1}{2}\left(\frac{\gamma_{e}t}{\hbar}\right)^{2}}\\
 & +\frac{2}{3\pi}\left(\sqrt{\frac{1}{f(t)^{2}}-1}+\left(2-\frac{1}{f(t)^{2}}\right)\arcsin(f(t))\right).
\end{align*}

For the heavy-hole as the central spin we have $C_{z}=\alpha B_{z},\;\;C_{x,y}=\beta B_{x,y}$
with $\alpha=\gamma_{h_{z}}$ $\beta=\gamma_{h_{p}}$ In this case
$\langle S_{z}\rangle$ is given according to Eq.~(\ref{eq:St})
by a sum of two rather complicated integrals. The second term of Eq.~(\ref{eq:St})
can be reduced into a one-dimensional (1D) integral which we than
calculate numerically 
\begin{equation}
\frac{\beta^{2}}{\left(\alpha^{2}-\beta^{2}\right)^{3/2}}\int_{\beta}^{\alpha}d\xi\,\frac{(\alpha^{2}-\xi^{2})(1-\sigma^{2}\xi^{2}t^{2})}{\xi\sqrt{\xi^{2}-\beta^{2}}}e^{-\frac{1}{2}\sigma^{2}\xi^{2}t^{2}}.\label{Hole1}
\end{equation}
The first term of Eq.~(\ref{eq:St}) is a more complicated 6D integral.
If we use the following shorthands 
\[
a_{0}=\sqrt{\left(\frac{\cos^{2}\theta}{\alpha^{2}}+\frac{\sin^{2}\theta}{\beta^{2}}\right)\left(\frac{\cos^{2}\theta'}{\alpha^{2}}+\frac{\sin^{2}\theta'}{\beta^{2}}\right)},
\]
\[
a_{1}=f(t)\left(\frac{\cos\theta\cos\theta'}{\alpha^{2}}+\frac{\sin\theta\sin\theta'}{\beta^{2}}\cos\varphi\right),
\]
\[
A_{0}=\frac{(1-f(t)^{2})^{3/2}}{4\pi^{2}\alpha^{2}\beta^{4}}\sin(2\theta)\sin(2\theta')
\]
then the 6D integral can be reduced into a 3D one
\begin{align}
\int_{0}^{2\pi}d\varphi\int_{0}^{\pi/2}d\theta\int_{0}^{\pi/2}d\theta'\,A_{0}\biggl[\frac{3a_{1}}{(a_{0}^{2}-a_{1}^{2})^{2}}\nonumber \\
+\frac{a_{0}^{2}+2a_{1}^{2}}{(a_{0}^{2}-a_{1}^{2})^{5/2}}\arcsin(a_{1}/a_{0})\biggr],\label{Hole2}
\end{align}

which we than calculate numerically. The function $f(t)$ is essentially
the Overhauser field time correlator. An appropriate model for the
evolution of the Overhauser field is required for its evaluation. 

By using $\vec{B}=\frac{\nu_{0}}{g_{e}\mu_{B}\sigma}\sum A_{i}|\psi_{env}(\vec{r}_{i})|^{2}\vec{I}_{i}$
one obtains: 
\begin{align*}
3\sigma^{2}f(t) & =\langle\vec{B}(0)\cdot\vec{B}(t)\rangle\\
= & \sum\left(\frac{\nu_{0}}{g_{e}\mu_{B}\sigma}A_{i}^{2}|\psi_{\mathrm{env}}(\vec{r}_{i})|^{2}\right)^{2}\langle\vec{I}_{i}(0)\cdot\vec{I}_{i}(t)\rangle.
\end{align*}
A particularly simple model assumes that the Overhauser field evolution
is dominated by the quadrupole interaction of the nuclear spins \citep{Sinitsyn2012,Bechtold2015}.
Though more complicated models exist as well \citep{Efros2002,Bechtold2015,Al-Hassanieh2006},
this model permits analytical solutions.

\begin{figure}
\begin{centering}
\includegraphics[width=1\columnwidth]{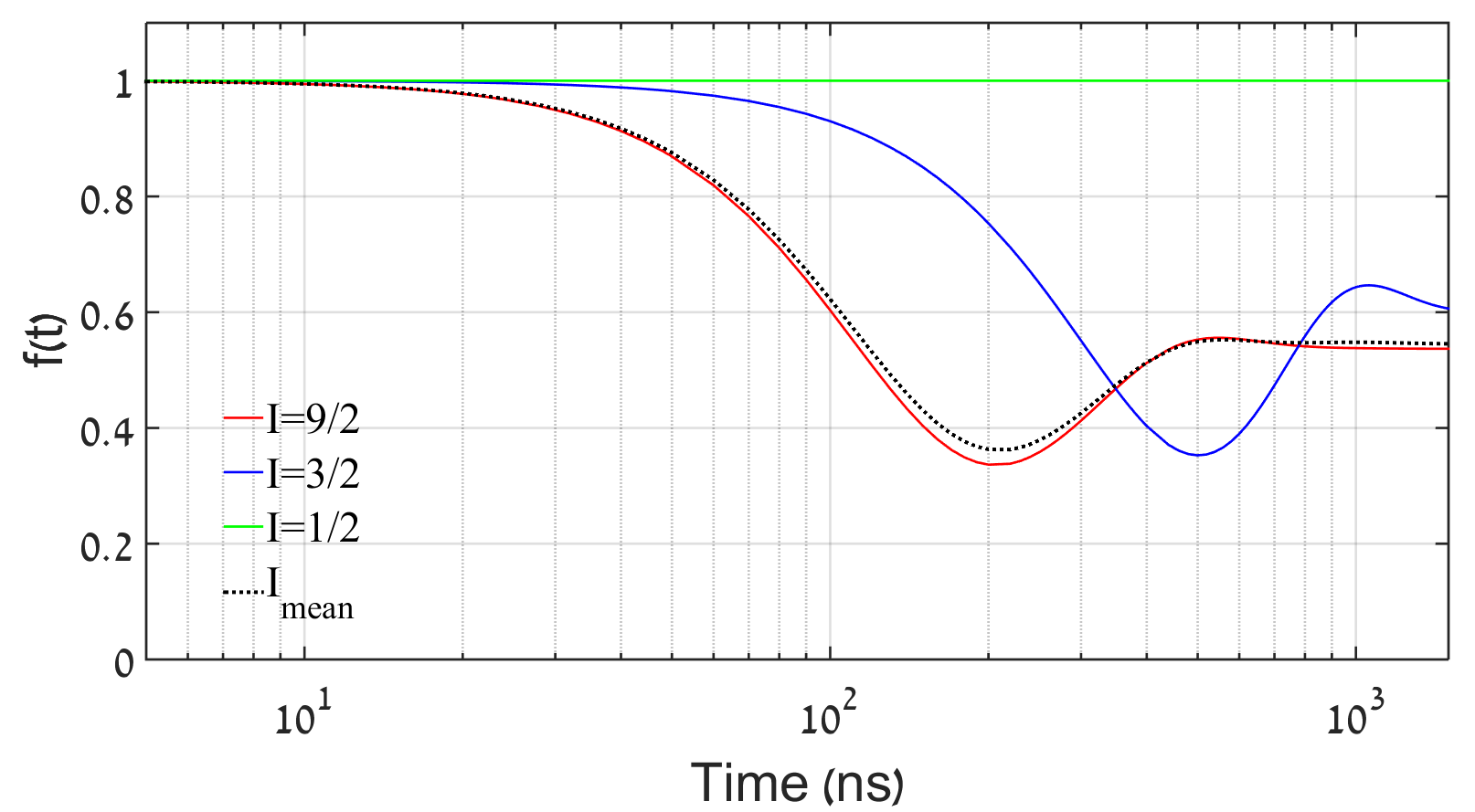}
\par\end{centering}
\caption{\label{fig:f(t)}The calculated normalized Overhauser-correlator $f_{I}(t)$
for the various types of nuclear spins which comprise the QD. $I_{\mathrm{mean}}$
is the mean value of the correlator taking into account the isotopic
abundances weighted by their squared magnetic moments.}
\end{figure}

Within this model each nuclear spin $\vec{I}_{k}=\vec{I}$ evolves
independently of the others by a Hamiltonian of the form $H_{Q}=V_{ij}I_{i}I_{j}$
with random $V_{ij}=V_{ji}$ which relates to the local electric field
gradients \citep{Sinitsyn2012} (EFG). We take the initial state of
the nuclear spin to be random and we average over the corresponding
wave function thereby obtaining 
\[
\langle\vec{I}(0)\cdot\vec{I}(t)\rangle\propto\Tr\left(\vec{I}\cdot e^{iH_{Q}t}\vec{I}e^{-iH_{Q}t}\right)
\]
As different nuclear spins have different EFG we obtain the Overhauser-correlator
$f(t)$ by averaging over the$V_{ij}$ terms. We take (as common in
random matrix theory) the elements of the symmetric matrix $V_{ij}$
to be independent Gaussian random variables of variance $\gamma_{Q}^{2}$.
Up to overall normalization we obtain 
\[
f(t)\propto\int dV\,e^{-TrV^{2}/(2\gamma_{Q}^{2})}\Tr\left(\vec{I}\cdot e^{iH_{Q}t}\vec{I}e^{-iH_{Q}t}\right).
\]
Noting that $V$ can be taken as a traceless tensor and in addition
using its polar decomposition reduce the above expression into a two-dimensional
integral which we express as 
\begin{align*}
f(t) & \propto\int dx_{1}dx_{2}dx_{3}\,\delta\left(\sum x_{i}\right)\prod_{i<j}|x_{i}-x_{j}|e^{\frac{-(x_{1}^{2}+x_{2}^{2}+X_{3}^{2})}{2\gamma_{Q}^{2}}}\\
 & \cdot\Tr\left(\vec{I}\cdot e^{i\sum x_{i}I_{i}^{2}t}\vec{I}e^{-i\sum x_{i}I_{i}^{2}t}\right).
\end{align*}
For $I=\frac{3}{2}$ we evaluate this expression and obtain: 
\[
f_{\frac{3}{2}}(t)\propto\int_{0}^{\infty}dx\,x^{4}\,e^{-x^{2}/(2\gamma_{Q}^{2})}(3+2\cos(\sqrt{6}xt))
\]
\begin{equation}
f_{\frac{3}{2}}(t)=\frac{3}{5}+\frac{2}{5}\left(1-2\left(\frac{\gamma_{Q}t}{\hbar}\right)^{2}+12\left(\frac{\gamma_{Q}t}{\hbar}\right)^{4}\right)e^{-3\left(\frac{\gamma_{Q}t}{\hbar}\right)^{2}}.\label{jygjy}
\end{equation}
 For higher values of the nuclear spin $I$, we calculated $f_{I}(t)$
numerically as a function of the dimensionless product $\gamma_{Q}t/\hbar$.
This gave qualitatively similar result to Eq.~(\ref{jygjy}) with
some modifications. Since our QD contains $I=\frac{3}{2}$ , $I=\frac{9}{2}$
and $I=\frac{1}{2}$ we averaged over these values using the relative
nuclear abundance multiplied by the squared nuclear moments as weights.
In Fig.~\ref{fig:f(t)} we display the normalized Overhauser-correlator
for various types of nuclear spins in the QD. For simplicity we assume
the same $\gamma_{Q}$ for all atom types. In practice the Indium
contribution dominates the average due to its large magnetic moment.

———

\section{Hyperfine interaction of the dark exciton.}

The spin projection ($S_{z}(t)$) of the DE strongly depends on the
electron-hole exchange interaction. 

We describe the DE qubit by its two spin states: $|\Uparrow\uparrow\rangle$
and $|\Downarrow\downarrow\rangle$ , with $J_{z}=+2$, and -2 respectively.
While the DE interaction with the z-component $B_{z}$ of the Overhauser
field is similar to that of a spin $\frac{1}{2}$ central spin (up
to multiplicative constant \citep{Bayer2002}), its interaction with
the $B_{x,y}$ components is very different. Strictly speaking, a
standard $\vec{B}\cdot\vec{J}$ Hamiltonian would have to act four
times in order to flip a $J_{z}=2$ state into $J_{Z}=-2$ state.
However, if one fully considers the electron-hole exchange interaction,
this is not the case. In the bright and dark excitons basis $\{|\Uparrow\downarrow\rangle,|\Downarrow\uparrow\rangle,|\Uparrow\uparrow\rangle,|\Downarrow\downarrow\rangle\}$
, the exchange interaction can be expressed as \citep{Bayer2002,Don2016}
\[
\frac{1}{2}\left(\begin{array}{cccc}
\Delta_{0} & \Delta_{1}^{*} & \Delta_{3} & \Delta_{4}\\
\Delta_{1} & \Delta_{0} & -\Delta_{4}^{*} & -\Delta_{3}^{*}\\
\Delta_{3}^{*} & -\Delta_{4} & -\Delta_{0} & \Delta_{2}^{*}\\
\Delta_{4}^{*} & -\Delta_{3} & \Delta_{2} & -\Delta_{0}
\end{array}\right),
\]
where $\Delta_{0}$ is the isotropic exchange interaction. It is a
real number, which defines the energy splitting between the DE and
BE eigenstates. It was measured to be $\Delta_{0}$=260~\textmu eV
for the QD under study. The term 
\[
\Delta_{1}=\delta_{1}\exp(i2\theta_{1})
\]
is the anisotropic long-range exchange interaction. Here $\delta_{1}$
is a positive number defining the magnitude of the bright exciton
(BE) fine structure splitting (FSS) \citep{Ivchenko2005}, and $\theta_{1}$
defines the directions of the two cross linearly polarized components
of the BE spectral lines with respect to the crystallographic directions
\citep{Bayer2002}. 
\[
\Delta_{2}=\delta_{2}\exp(i2\theta_{2})
\]
 describes the FSS of the dark exciton. Here $\delta_{2}$ and $\theta_{2}$
are real numbers mainly given by the short range anisotropic exchange
interaction. 
\[
\Delta_{3}=\delta_{3}\exp(i2\theta_{3})
\]
 and
\[
\Delta_{4}=\delta_{4}\exp(i2\theta_{4})
\]
are also long-range exchange interactions that couple between the
DE and BE states. 

Strictly speaking, for a $C_{3v}$ symmetrical QD, $\delta_{1}$,
$\delta_{2}$, $\delta_{3}$ and $\delta_{4}$ are all expected to
vanish \citep{Dupertuis2011}. To within our experimental uncertainty,
we found it to be true only for $\delta_{2}(<0.1\,\text{\textmu eV}$),
since it results from the short range exchange interaction and therefore
affected mainly by the symmetry of the QD's unit cells \citep{Bayer2002}.
Structural deviations of the QD from symmetry such as composition
fluctuations, or faceting, destroy the QD long range symmetry, without
affecting its unit cell symmetry. Therefore, they will result in finite
$\delta_{1}$ , $\delta_{3}$ and $\delta_{4}$. Indeed, we measured
$\delta_{1}=18\,\text{\textmu eV}$ by polarization sensitive spectroscopy,
and estimated $\delta_{3}\backsimeq\delta_{4}\backsimeq15\,\text{\textmu eV}$
by measuring the DE radiative lifetime, and verifying the fact that
the DE weak absorption line was linearly polarized in-plane \citep{Schwartz2015}
. 

Since $|\Delta_{3}|=|\Delta_{4}|$$\neq0$ , these terms induce coupling
between the DE and BE states. We define $\left(\Delta_{3}+\Delta_{4}\right)/2\triangleq\Delta_{\mathrm{DB}}$,
and since $|\Delta_{\mathrm{DB}}|\ll\Delta_{0}$ , the modified DE
eigenstates remain almost degenerate such that the symmetric and anti-symmetric
spin combinations are expressed as 
\begin{align*}
\ket{\mathrm{DE}_{AS}} & =N_{AS}\left[\frac{\ket{\Uparrow\uparrow}-\ket{\Downarrow\downarrow}}{\sqrt{2}}-\frac{\Delta_{\mathrm{DB}}}{\Delta_{0}}\frac{\ket{\Uparrow\downarrow}+\ket{\Downarrow\uparrow}}{\sqrt{2}}\right]\\
\ket{\mathrm{DE}_{S}} & =\frac{\ket{\Uparrow\uparrow}+\ket{\Downarrow\downarrow}}{\sqrt{2}},
\end{align*}

where $N_{AS}\sim1$ is a normalization constant. This also agrees
with the experimental observation that the DE has only one weak optically
active eigenstate, which is linearly polarized like the symmetric
BE eigenstate \citep{zielinsky2014,Schmidgall_2017,Don2016}. The
mixing term is sufficient to provide a nuclear field dependent flipping
of either the heavy hole or the electron in order to change the DE
state from the $\ket{\Uparrow\uparrow}$ to the $\ket{\Downarrow\downarrow}$
or vice versa. Hence, the interaction is linear in the nuclear magnetic
field and the DE Hamiltonian takes the form $H=\frac{1}{2}\vec{C}^{\mathrm{(DE)}}\cdot\vec{\sigma}$
with 
\begin{align*}
C_{x,y}^{\mathrm{(DE)}} & =\frac{2\im[\Delta_{\mathrm{DB}}]}{\Delta_{0}}\left(C_{x,y}^{(e)}+C_{x,y}^{(h)}\right),\\
C_{z}^{\mathrm{(DE)}} & =\left(C_{z}^{(e)}+C_{z}^{(h)}\right).
\end{align*}
 If we express $\vec{C}_{e},\vec{C}_{h}$ as earlier in terms of the
same dimensionless $\vec{B}$, we conclude 
\begin{align}
\gamma_{\textrm{\textrm{D}E}{}_{p}} & =\frac{2\im[\Delta_{\mathrm{DB}}]}{\Delta_{0}}\left(\gamma_{e}+\gamma_{h_{p}}\right),\label{eq:gamma_DE_p}\\
\gamma_{\textrm{DE}_{z}} & =\gamma_{e}+\gamma_{h_{z}}=\gamma_{e}-|\gamma_{h_{z}}|,\label{eq:gamma_DE_z}
\end{align}
where we used the fact that $\gamma_{h_{z}}<0$ \citep{Witek2011}.

$\im[\Delta_{\mathrm{DB}}]\leqq\delta_{3}\approx15\,\text{\textmu eV}$
provides an estimate for $\gamma_{\textrm{\textrm{D}E}{}_{p}}$ (see
Table.~\ref{tab:Interaction-energies}), and we note here that the
fields $\vec{C}_{e}$ and $\vec{C}_{h}$ experienced by the electron
and by the heavy-hole, respectively, may not be in perfect correlation
\citep{Fischer2008}. This is expected to reduce their interference
effects, making $\gamma_{\textrm{DE}_{z}}$ slightly larger and $\gamma_{\textrm{\textrm{D}E}{}_{p}}$
slightly smaller than the above estimations.

The DE Hamiltonian as explained above is linear in B and anisotropic,
much like the one for the heavy-hole spin. Consequently $\langle S_{z}(t)\rangle$
is derived in a similar way to that of the heavy-hole spin in Eq.~(\ref{Hole1})
and Eq.~(\ref{Hole2}) by replacing $\alpha=\gamma_{\textrm{DE}_{z}}$
and $\beta=\gamma_{\textrm{\textrm{D}E}{}_{p}}$.


\bibliography{spin_relaxation_dynamics}

\end{document}